\documentstyle [psfig,aps,prl,eqsecnum,preprint,tighten]{revtex}
\begin{document}   
\input psfig.sty

\title{Edge state transmission, duality relation and its implication to
measurements} 
\author{Shanhui Xiong} \address{International 
Center for Theoretical Physics, 34100 Trieste, Italy}
\date{\today}
\maketitle
\newcommand{\und}{\underline}
\newcommand{\ov}{\overline}
\newcommand{\be}{\begin{equation}}
\newcommand{\ee}{\end{equation}}
\newcommand{\bea}{\begin{eqnarray}}
\newcommand{\eea}{\end{eqnarray}}
\newcommand{\br}{{\bf r }}
\newcommand{\bA}{{\bf A }}
\newcommand{\brp}{{\bf r}^{\prime} }
\newcommand{\bna}{{\nabla}}
\newcommand{\DD}{\stackrel{\leftrightarrow}{D}}
\newcommand{\DDs}{\stackrel{\leftrightarrow}{D^*}}
\newcommand{\non}{\nonumber}
\newcommand{\dl}{\delta}
\newcommand{\al}{\alpha}
\newcommand{\om}{\omega}
\newcommand{\gm}{\gamma}
\newcommand{\la}{\lambda}
\newcommand{\wc}{\omega_c}
\newcommand{\Gprrp}{G^+(\br,\brp)}
\newcommand{\Gmrrp}{G^-(\br,\brp)}
\newcommand{\Gprpr}{G^+(\brp,\br)}
\newcommand{\Gmrpr}{G^-(\brp,\br)}
\newcommand{\rarrow}{\rightarrow}
\newcommand{\gpm}{G^{\pm}}
\newcommand{\gnpm}{G_0^{\pm}}
\newcommand{\Gb}{\ov{G}}
\newcommand{\Gbp}{\ov{G}^+}
\newcommand{\Gbm}{\ov{G}^-}
\newcommand{\Gbrrp}{\ov{G}(\br,\brp;E)}
\newcommand{\Gnrrp}{G_0(\br,\brp;E)}
\newcommand{\dlA}{\Delta \bA}
\newcommand{\sx}{\sigma_{xx}}
\newcommand{\sy}{\sigma_{xy}}
\newcommand{\ea}{\end{array}}
\newcommand{\ba}{\begin{array}}
\newcommand{\fba}{\ov{\varphi}}
\newcommand{\zd}{z^{\dagger}}

\begin{abstract}

The duality in the Chalker-Coddington network model is examined. We are
able to write down a duality relation for the edge state transmission 
coefficient, but only for a specific symmetric 
Hall geometry. Looking for broader implication of the duality, 
we calculate
the transmission coefficient $T$ in terms of 
the conductivity $\sx$ and $\sy$ in the diffusive limit. The edge state
scattering problem is reduced to solving the diffusion equation with two
boundary conditions
$(\partial_y-\textstyle{\frac{\sy}{\sx}}\partial_x)\phi=0$ and
$[\partial_x+\textstyle{\frac{\sy-\sy^{lead}}{\sx}} \partial_y]\phi=0$.
We find that the resistances  in the geometry considered are not
necessarily  measures of the resistivity 
and $\rho_{xx}=\textstyle{\frac{L}{W}\frac{R}{T}\frac{h}{e^2}}$ ($R=1-T$)
holds 
only when $\rho_{xy}$ is quantized.  We conclude
that duality alone is not sufficient to explain the
experimental findings of Shahar et al and 
that Landauer-Buttiker argument does not render the additional condition,
contrary to previous expectation. 
  
\end{abstract}

\section{Introduction}

We consider the transport properties of a two-dimensional (2D) 
disordered strip connected to two disorder-free  leads, as shown in figure 1. 
In the transverse direction, the
system is restricted to a finite width of $W$ by confinement potentials.
The entire system
is subject to a perpendicular magnetic field. In the leads the edge
states are the only current carrying states. Therefore the problem
can also be viewed as that of edge states 
scattering through a disordered region. From  the viewpoint of the
Landauer-Buttiker  formula\cite{Landauer,Buttiker},  the DC transport 
properties  are determined by
the scattering matrix at the Fermi energy.   It has been shown that
the total transmission or the total back scattering of edge states
leads to the quantization of Hall
conductance\cite{Halperin,Streda0,Jain,Buttiker1};
from the bulk point of view  the quantization requires the localization of
the bulk states.  The equivalence between the  edge state 
description and the  bulk point of view has generated interesting
discussions\cite{Halperin,Streda0,Jain,Buttiker1,Pruisken,Aaron}. However,
there
has been no quantitative analysis
as to how the edge state transmission coefficient relates to the bulk
conductivity or resistivity.  It has been shown\cite{Streda0,Jain} that 
the longitudinal resistance $R_{xx}$ of the above system is
$$R_{xx}=\frac{R}{T}\frac{h}{e^2}, $$
where $T$ and $R$ are transmission and reflection coefficient of a single
edge
state. The following relation 
$$
\rho_{xx}=\frac{L}{W}\frac{R}{T}\frac{h}{e^2}
$$
has also been used in a number of articles\cite{Jain,Aaron}, where
$\rho_{xx}$ is the longitudinal resistivity, $W$ and $L$ is the width and
length of the sample respectively. 
The latter relation is based on the assumption that the resistance is
a measure of the resistivity. This assumption is not justified.

Our work is also motivated by the recent experimental and subsequent
theoretical work by Shahar et al\cite{Shahar1,Shahar2} concerning the
duality between phases on either sides of the quantum Hall transitions.
Using the Chalker-Coddington network model\cite{Chalker}, we are
able to obtain a previously speculated duality relation\cite{Shahar2}
for the transmission coefficient and the longitudinal resistance in the
geometry specified above.  To decide whether or not the duality relation
explains the experiment, which was done in a different geometry,   one
again needs to resolve the relation between resistance and 
resistivity.

This   paper  addresses two issues.  First we are concerned with the
duality in
Chalker-Coddington model and  how this duality
manifests itself in  resistance measurement.  
Secondly we  render a microscopic calculation for the edge state
transmission coefficient in the
diffusive limit in terms of the bulk parameters,
the longitudinal and Hall conductivity $\sx$ and $\sy$. 
We find a non-trivial relation between resistance
and resistivity. Although the calculation helps to put a restriction to the
implication of the duality, it resolves  an independent issue of its own.

Section \ II serves as a brief review
of the quantum linear response theory. There we set up the starting point
for the microscopic calculations.   We emphasize that for finite-sized
systems with phase coherence, all measured quantities are conductances or 
resistances and  are in principle sensitive to the way the
measurement is set up.  We make contrast  between 
the bulk current density  and  the  
Landauer-Buttiker scattering point of view point. In following both
approaches in later calculations, we demonstrate  their equivalence.

In section \ III we express the Hall and longitudinal resistance  
in terms of the transmission coefficient using the Landauer-Buttiker
formula.  Our derivation makes explicit the involvement of probes and
leads in the measurement. In section\ IV, we proceed to discuss the
duality in 
the discrete Chalker-Coddington
model, which leads to an inverse relation for the
longitudinal resistance, but only for  samples with
reflection symmetry.  We also look for the implication of the duality
in the continuous limit. There the duality is between the
phase with its bare Hall conductivity at $\sy^0$ and the phase  
at $n-\sy^0$, where $n$ is a integer. (Note: from 
now on we use
$e^2/h$ as the unit for conductivity and $h/e^2$ for resistivity.)
The duality relation in the continuous limit translates to a relation
between the renormalized conductivity: $\sy(\sy^0)+\sy(n-\sy^0)=n$ (here 
$\sy(\sy^0)$ denotes the renormalized $\sy$ as a function its bare value 
$\sy^0$). We 
show that this duality relation alone is not enough to explain the
experimental claim of \cite{Shahar1}. To do so  an additional
constraint
between $\sx$ and $\sy$ is required.  

In section \ V we calculate the transmission
coefficient and the resistances in the perturbative limit ($\sx>1$). 
We show that  the ideal leads affect the outcome of the measurement by 
imposing a boundary condition on the electro-chemical potential
and the Hall resistance is thus artificially fixed to a quantized value.
We find that only under the condition that the Hall resistivity is
quantized the resistance is
proportional to resistivity. The
edge state scattering problem in the diffusive limit is reduced to a
special boundary problem. Its analytical solution by conformal mapping
is discussed in section \ VI. We conclude  with a
discussion on  the  missing connections between the duality found in
experiment and that in existing models for the quantum Hall effects.

\section{Two forms of the linear response theory}

The quantum mechanical linear response theory for non-interacting
electron gas can be put in two  forms. In one form one writes 
the local current density as a functional of the external field:
\be
j_{\mu}(\br)=\int d\brp
\sigma_{\mu\nu}(\br,\brp)E_{\nu}(\brp),\label{bilocal}
\ee
where $\sigma_{\mu\nu}(\br,\brp)$ is the bilocal conductivity and 
can be expressed in terms of the single particle Green's function
$G^{\pm}(E)=1/(E-\hat{H}\pm i\eta)$ where $H$ is the single particle
Hamiltonian. (For detailed form of $\sigma(\br,\brp)$ see
\cite{Baranger}). 
The relation between  the current and  the external field is generally
non-local and in the presence of magnetic field
$\sigma_{\mu\nu}(\br,\brp)$ contains not only Fermi surface
contribution, but also contributions from all energies below the Fermi
energy. 

Due to a set of current  conservation constraints \cite{Baranger}, 
the total current in a lead (say, the $i$th lead) to linear order depends
only
on the voltages in the leads, the $V_j$s:
\be
I_i=g_{ij}V_j.
\ee
where the  conductance coefficients, the
$g_{ij}$s, are surface integrals of $\sigma(\br,\brp)$ in the $i$th and
$j$th lead:
\be
g_{ij}=-\int \int d{\bf S}_i
\cdot \sigma(\br,\br') \cdot d{\bf S}_j. 
\ee   
Although the off-Fermi-surface terms in $\sigma(\br,\brp)$ contributes to
the local current response, they give zero net
contribution upon surface integral\cite{Baranger}. As a result the
$g_{ij}$s  can be written in terms of the scattering matrix at the Fermi
energy. One arrives at the Landauer-Buttiker
formula\cite{Landauer,Buttiker}
\be
g_{ij}=\frac{e^2}{h}T_{ij},\;\mbox{for}\; i\neq j;\;\;\;
g_{ii}=\frac{e^2}{h}\left[T_{ii}- N \right]\label{Landauer}
\ee
where $T_{ij}$ is the total transmission coefficient from the $i$th to the
$j$th lead,
$$
T_{ij}=Tr\left\{t_{ij} t^{\dagger}_{ij}\right\};
$$
$t_{ij}$ is  the scattering matrix
between the states of the $i$th and the $j$th probe and $N$ is total
number of scattering states (or total number of edge states in the
presence case) at the Fermi-level. The 
above formula best illustrates the no-local aspect of quantum
transport  and  has been instrumental to our understanding
of  Anderson localization and to the formulation of random matrix
theory for  quasi one-dimensional disorder
systems. The DC conductance $g_{ij}$ is
well defined for finite-sized disordered region (embedded in an infinite
open system), therefore are good candidates for scaling analysis;
moreover, they are directly measurable in experiments. However, the
formula  
is rarely  used  for microscopic calculations, because it
introduces confinements and leads which can be cumbersome to
address theoretically, particularly in the presence of magnetic field.  
Only recently it became understood that the presence of confinement
potential at the edges changes the boundary 
condition for diffusion from $(\partial_n)\phi
=0$ to $(\partial_n +
\textstyle{\frac{\sy}{\sx}} \partial_t)\rho=0$ where $t$ is the 
tangential and $n$ is the normal direction of the edge\cite{RKM}. The
boundary condition imposed by the perfect
2D leads will be discussed in this paper.

In practice, the conductivity, a concept from classical physics, is still
widely used, although for quantum mechanical system it no longer bears the 
local interpretation. It can be defined as the ratio between the average
current density and the average field. Caution has to be exercised for
finite-sized closed systems, where the 
DC dissipative conductivity $\sx$ is zero  due to the discreteness of 
energy levels. The DC  $\sx$ is usually 
asymptotically defined from AC conductivity $\sx(\omega)$ 
by taking the system size to infinite before taking  the frequency
($\omega$) of the external field to
zero.  Experimentally  most of measurements of  $\sx$ and $\sy$
are deduced from longitudinal and Hall conductance(resistance) by extending 
the following classical relation to quantum mechanical systems:
\be
g_{xx}=\sx\frac{W}{L}, \;\; g_{xy}=\sy, \label{scaling}
\ee
where $g_{xx}$ and $g_{xy}$ are combinations of the transmission
coefficients. 
The ``conductivities" thus defined are in fact conductances. 
 The conductance or the transmission coefficients  depend on a number of
factors: the random scattering potential in the disordered region,  the
confinement potential, the property of the leads and even that of the
thermal reservoirs. Result of a measurement  is not independent of the way
in which the measurement is done.
Without checking that the measurement results are robust 
against size and geometry variations, one can not be sure that the 
conductivity (or resistivity) are obtained.
As we will show in section\ V, even in the perturbative regimes
where quantum corrections are small, $R_{xx}$ and
$R_{xy}$ exhibit the above Ohmic type of 
relations  only under certain condition. 
For the critical regimes  between the quantum
Hall plateaus, experimental and 
numerical studies  for different
geometries have so far come up with a range of  values for
the critical value of $\sx$\cite{Wang}, indicating the need for more
careful study of finite-size scaling under different boundary conditions.  
We would like to address this point in our future study.

\section{Landauer-Buttiker formula for resistance}

In this section, we express the longitudinal
and Hall resistance of our system  in terms of
the total transmission coefficient. These Landauer-Buttiker type of
expressions have been derived before\cite{Streda0,Jain} and  recently
have been
evoked by Shahar et al as a possible vehicle for the understanding of
certain
duality relation in the quantum Hall effect. Our derivation stress
peculiarity of the measurement involved.

The system as it is shown in figure 1 can not be used to extract
both the Hall and longitudinal resistance. To do so while preserving the
simplicity of the S-matrix, one can attach 
additional voltage probes outside the scattering
region. Suppose we attach 4 voltage probes and make a 6-probe  Hall
bar, as
shown in figure 2a. The edge states outside the scattering region  go from
one probe 
to the next with probability $1$ and  all measurement result are related
to one
parameter, the transmission coefficient $T=Tr \{t t^{\dagger}\}$, where
$t$ is the $N\times N$ transmission matrix of the right-going edge states
over the
disordered region. The
transmission coefficient of the left-going edge
states  is also $T$ by the unitary requirement of the S-matrix.  
The only non-zero elements among the
$T_{ij}$s are: $T_{21}=T_{43}=T_{54}=T_{16}=N$, $T_{32}=T_{65}=T$,
$T_{62}=T_{35}=R$, where $R=N-T$.  Passing a total
current $I$ in the x-direction, between probe 1 and 4, the resulting
voltages can be solved (up to a constant shift) using the
Landauer-Buttiker formula (\ref{Landauer}), bearing in
mind that the total current in the voltage probes is fixed to be zero.
The Hall resistance $R_{xy}=(V_3-V_5)/I=(V_2-V_6)/I$  
in this particular
setup   is totally dictated by the property of the ideal leads
since  the voltage probes are across the  leads. In the leads,  
the Fermi energy is pinned to  its gap of bulk density of states. In
such case, it is easy to show that the integrated current is equal to $N $
multiplied by
the voltage difference of the two  edges. Therefore, 
\be
R_{xy}=\frac{1}{N}.\label{Rxy}
\ee
The longitudinal voltage drop can be shown to be\cite{Streda0,Jain} 
\be
R_{xx}=\frac{V_3-V_2}{I}=\frac{V_5-V_6}{I}
=\frac{1}{N}\frac{R}{T}.\label{Rxx} 
\ee

We emphasize that while the above expressions hold for every realization
of the disorder, they are not expected to hold for other geometries. As
we have mentioned, the quantization of $R_{xy}$ is a direct consequence of
the measurement arrangement; it does not imply that $\rho_{xy}$ is also
quantized. Note that the quantization of $\rho_{xy}$ implies that 
$\sx^2+\sy^2=N\sy$. Were it to be true, this would give, 
for critical  $\sy=\sy^c=N-1/2$, the critical
$\sx=\sx^c=\textstyle{\frac{\sqrt{2N-1}}{2}}$, which contradicts with the
current
consensus that all integer quantum Hall
transitions are the same and that $\sx^c$ should be independent of $N$.
This is one more reason  that $R_{xx}$ should not be taken as $\sx$ for
$N>1$. (This simple measurement setup is not able to detect whether N Landau
levels are
coupled by disorder or decoupled from one another
in the scattering region, since it is  only sensitive to the trace of the
transmission matrix. Equilibrium among edges states are re-enforced by the
thermal reservoirs attached to the probes.)

\section{Duality relation for the edge transmission coefficient}

Recently, in studying of the critical transitions between the quantum
Hall plateaus, Shahar et al\cite{Shahar1,Shahar2} find that the
longitudinal I-V curves near the
critical magnetic field $B_c$ are non-linear and demonstrate a certain
reflection symmetry with respect to the
linear line at $B_c$; more precisely, the longitudinal voltage  and
current appear to reverse roles at two filling fractions, $\nu$ and
$\nu^d$, on either sides of the quantum Hall transition:
\be
\{V_x(\nu^d),I_x(\nu^d)\}=\{I_x(\nu),V_x(\nu)\},\label{duality1}
\ee
with $\nu$ and $\nu^d$ satisfying a definite relation suggestive of
charge-flux symmetry in the bosonic view of the quantum Hall 
effect\cite{Shahar1,Shahar2}. The authors point out that duality
between charge and flux in the effective bosonic action can be the
explanation of the observed relation (\ref{duality1}), however there is
one
catch to this scenario-- it requires  the extra condition that the bosonic
 Hall resistivity remains zero across the phase transition, to which
there has been no satisfying explanation. As an alternative the 
same authors also suggest  a fermionic scenario, appealing to the
Landauer-Buttiker expression similar to (\ref{Rxy}) and (\ref{Rxx})
(their version is amendable to linear response). Noticing that for
$N=1$, $R_{xx}$ goes to its inverse as $T \rightarrow 1-T=R$,  
Shahar et al propose that (\ref{duality1}) can be explained within the
Landauer-Buttiker framework  if the following
is true:
\be
T(E_c+\Delta E)=1-T(E_c-\Delta E).\label{Ansatz}
\ee
We show  that the above relation can be the consequence of certain
duality embedded in the Chalker-Coddington model for the quantum Hall
effect.
However, one can only write down the duality relation for
the transmission coefficient of a symmetric sample with $L=W$.  

The Chalker-Coddington model\cite{Chalker}   
consists of a lattice  of directed links and scattering nodes (see figure 
3), representing the semi-classical orbits along the  equipotential
contours of the random potential and the tunneling among these orbits at
the nodes.
Each node is described by a
$2\times 2$
scattering matrix with random phases and a fixed
probability $T_0$ to scatter to the right and $1-T_0$ to the left.
$T_0$ is a function of the Fermi-energy. 
In figure 3, we  show one such network coupled to two edge states. 
At fixed Fermi energy, there
are two equivalent ways to view the network: as one built out of
clockwise guiding center orbits (the white squares)  with probability
$1-T_0$ to tunnel to 
the neighboring orbits, or equivalently, 
of counter-clockwise orbits (black squares) with tunneling probability
$T_0$. (At $T_0=1$, the network breaks down
to decoupled clockwise orbits and two edges state at the top
and bottom; at $T_0=0$, it breaks down to decoupled counter-clockwise 
orbits and two edge states at the left and right entrance of the
sample.)   The two states at energies related by $T(E'_f)=1-T_0(E_f)$
are dual in the sense that one ensemble can be mapped onto the other, if
the system is infinite,  by reflection with respect to any of the discrete
ridges in the direction of $\hat{x}\pm \hat{y}$. If the system is finite   
and of arbitrary shape,  the above symmetry is broken by the boundary.
However, if the system is a square, two of the  reflection axis survive.
In this case the reflection along the diagonal brings the white squares
onto the back square, and at the boundary, transmission channel becomes
the reflection channel. 
Among many  relations one can write down between the two states (or
two phases), one is the following 
\be
\langle
T(1-T_0)\rangle=\langle R(T_0)\rangle.
\ee 
(In fact the distribution of $T$ at $1-T_0$ is identical to
that of $R$ at $T_0$.) The self-dual point is 
apparently  $T_0=1-T_0=T_c=1/2$. Chalker and Coddington has made use of
this fact in locating the critical point of the model. 
Linearizing the function $T_0=T_0(E_f)$ near $T_c$ ($E_c$), $
E-E_c\approx (T_0-T_c)/T'_0(E_c)$,
the above gives equation (\ref{Ansatz}) and subsequently 
\be
R_{xx}(E_c+\Delta E)=\frac{1}{R_{xx}(E_c-\Delta E)}. \label{duality2}
\ee
for $(T_0-T_c)/T_c\ll 1$. (The above is rigorously true if the random
potential distribution  is symmetric with respect to positive and negative
potentials. In that case it is appropriate to assume
$T_0(E_c+\Delta E)=1-T_0(E_c-\Delta E)$. In this case there is no more
need for the expansion.) If experiment is done in
the geometry we consider, relation (\ref{duality2}) will result 
as a simple consequence of the duality between the phase at $E_f=E_f(T_0)$
and
the phase at $E_f'=E_f(1-T_0)$. 

The experiment by Shahar et al
was done in  a Hall bar geometry with the voltage probes  placed in the
interior of the sample as shown in figure 2b. The S-matrix for the
realistic Hall bar is significantly more complicated. Moreover 
the argument leading up to relation (\ref{duality2}) requires 
the reflection symmetry with respect to the two diagonal axis. There is no
apparent reason why it should apply to a realistic Hall bar geometry with
 no such reflection symmetry.  To find the possible connection between 
the  experiment and  the duality in
the microscopic non-interacting models for the quantum Hall effect,
one has   to find out 1) whether or not  a more general duality 
relation can be written down which is independent of
geometry and boundary conditions 2) whether or not under certain 
conditions resistance measured  in complicated geometries exhibit
tractable scaling laws with some universal coefficients and  exponents.
  
The relation we wrote down for the discrete lattice model is somewhat
artificial, since the network  is an idealization of the
mutually tunneling guiding orbits of arbitrary size and shape. 
It is then sensible to look for implication of duality in the continuous
theories. The network model
has been mapped onto the non-linear sigma model
with a topological term\cite{RLZ}. The latter was shown by
Pruisken et al to exhibit the appropriate asymptotic scaling property
required of the quantum Hall transitions\cite{Pruisken}.  If 
the
network model is coarse-grained at length scale much larger than
the lattice spacing, the clockwise and
counter-clockwise orbits  are lost, i. e. one can not tell the difference
of the $T_0$ and $1-T_0$ state in
the bulk, however differences does show up  at the
boundaries.  Consequently, the bulk characteristic of the two phases, i,
e. the  diffusion constant or  $\sx^0$ are the same, while
their bare Hall conductivity $\sy^0$ differ by one quanta. (It has been
shown that the bare conductivity of the network model 
are $\sx^0=\textstyle{\frac{T_0(1-T_0)}{T_0^2+(1-T_0)^2}}$ and 
$\sy^0=\textstyle{\frac{T_0^2}{T_0^2+(1-T_0)^2}}$\cite{Aaron,XRS}). 
Since reflection is the operation that maps the $T_0$ phase to the
$1-T_0$ phase, we consider  how the non-linear sigma model transforms 
as  under one reflection operation
($x\rightarrow
-x$ or $y\rightarrow -y$). The topological term, with $\sy^0$
as its coefficient, changes sign while the 
$\sx^0$ term remains the same.
Since the theory is periodic in
$\sy^0$, the corresponding dual phases are that parameterized by $\sy^0$
and $n-\sy^0$, with $n=\ldots,-2,-1,0,1,2,\ldots$. (Note: $\sy^0$ and
$n-\sy^0$ correspond to two filling factors or two Fermi-energies). Using
the renormalization equations given by Pruisken et al\cite{Pruisken}, one
can verify the following relation between the renormalized $\sx$, $\sy$
\bea
\sx(\sy^0)=\sx(n-\sy^0)\non\\
\sy(\sy^0)=n-\sy(n-\sy^0),\label{duality3}
\eea
if $\sx^0(\sy^0)=\sx^0(n-\sy^0)$. The above does not necessarily
lead to the desired relation $$\rho_{xx}(\sy^0)=\rho^{-1}_{xx}(n-\sy^0).$$
To
obtain the
above,  one more constraint is required between $\sx$ and $\sy$ at the
same $\sy^0$ (filling factor, Fermi energy). 
For example, the above
relation is satisfied if $\sy$ and $\sx$ obey (\ref{duality3}), as well as
the semi-circle relation 
\be
\sx^2+(\sy-1/2)^2=1/4,\label{semicircle}
\ee 
which is equivalent to 
\be
\rho_{xy}=1.\label{constant}
\ee
(Up to this pint, other form of the constraint is also possible. See later
discussions on this issue.) Therefore,  in both the fermionic and the
bosonic picture
for
the quantum Hall effect duality alone is not sufficient to explain the
experimental result(\ref{duality1}). Both approaches require one more
constraint: constant or vanishing Hall resistivity. As we explained before
and will
make more clear in the calculation to follow, the quantization
of Hall resistance in our particular setup does not necessarily imply the
quantization of Hall resistivity.
It has been shown that equation (\ref{semicircle}) or (\ref{constant}) 
hold
for
a classical version of the network model with no phase
interference\cite{Kucera,XRS,Shimshoni}. There are also numerical evidence that 
it holds for the original quantum version of the network
model\cite{Ruzin}.
Theoretically the reason of the constraint is not clear.

\section{Resistance in terms of the bare conductivity, perturbative
considerations}

We next calculate the transmission coefficient, $R_{xx}$ and $R_{xy}$  in
our  geometry in terms of $\sx$ and $\sy$ in the perturbative
limit. (The Chalker-Coddington model for one Landau level has very small
bare conductivity $\sx^0 < 1$, therefore it can not be treated
perturbatively.) We perform our calculation  for high Landau levels
($N>1$) and for the  short-ranged random potential model, of which the
bare  conductivity $\sx^0$ can be large. The calculation serves to
strengthen our view regarding duality in quantum Hall systems, it also
has an interest of its own, since the problem of edge state transmission
in the multi-scattering, diffusive regime has not been analyzed before.
Our treatment is at most phenomenological. Its rationale is as
following.  In our previous work\cite{XRS}, we have calculated the bilocal
conductivity $\sigma_{\mu\nu}(\br,\brp)$ to leading order in $1/\sx^0$. We
find
that to leading order the non-local relation between current $\bf{j}(\br)$
and external
field ${\bf E}(\br)$ can be decoupled and 
reduced to the familiar Drude
equation (with some modification to be specified later) 
\be
j_{\mu}(\br)=\sigma^{0}_{\mu\nu}\cal{E}_{\nu}(\br), \label{classical}
\ee 
where $\cal{E}=\bf {E}+\nabla \mu/e$ is the electro motive field ($\mu$ 
is the local chemical potential).  In the same work we also give
the generating function from which one obtains the full quantum mechanical
conductance.  The effect of the edges can be treated within  the non-linear
sigma model, with $\sx^0$ and $\sy^0$ as the
coupling constants and the presence of  the topological term 
alters the boundary condition for diffusion. In treating the quantum
mechanical system as though it was classical, we are assuming that the
field theory is renormalizable,
i. e., all the quantum corrections to conductance can be accounted for by
replacing the bare coupling constants with the renormalized reversion. 
The assumption  so far has met no contradiction at perturbative
level\cite{XRS}. 
The new ingredient is the discussion on  the boundary condition imposed by
the 2D perfect leads. We show that the computation of transmission
coefficient and of current
distribution  reduce to solving the Laplace equation with identical
boundary conditions.

In reference\cite{XRS} it has been shown  using the self-consistent
Born approximation(SCBA)\cite{Ando}, which gives the leading order in
$1/\sx^0$, 
that the bilocal conductivity tensor is of the following
form:
\bea
\langle\sigma_{\mu\nu}(\br,\brp)\rangle&=& 
\left[\sx^0\dl_{\mu\nu}+(\sy^{I,0}+\sy^{II,0})\epsilon_{\mu\nu}\right]
\dl(\br-\brp)\non\\
&&\mbox{}-\frac{1}{\sx^0}\left[\sx^0\partial_\mu
+\sy^{I,0}\epsilon_{\mu\mu'}\partial_{\mu'}+\sy^{II,0}\dl_{\mu
x}(\dl(y-W)-
\dl(y))\right]\non\\
&&\mbox{}\times\left[
\sx^0\partial'_{\nu}
-\sy^{I,0}\epsilon_{\nu\nu'}\partial'_{\nu'}-\sy^{II,0}\dl_{\nu x}(\dl(y'-
W)-\dl(y'))\right] d(\br,\brp)\non\\
&&+O(1/\sx^0, l/L),      \label{smunuSCBA}
\eea
where $d(\br,\brp)$ is the diffusion propagator satisfying the 
$-\nabla^2 d(\br,\brp)=\dl(\br,\brp)$, $\sx^0$, $\sy^{I,0}$ and
$\sy^{II,0}$ are the SCBA version of the Streda 
conductivities\cite{Streda,Pruisken,XRS}. The higher gradient terms are
of higher order in $l/L$, where $l$ is the mean free path and $L$ is the
system length.
The physics implied by the   leading order expression 
is  simple. The first term, the contact term, arises, in terms of 
the Drude's picture for conduction,  from 
electrons accelerating in the combined  external  field 
${\bf E}-e {\bf v}\times {\bf B} $. The second  term is
the diffusion term and it arises from the charge density fluctuation 
$\delta n_e$ in response to the external potential, which is 
 long-ranged. A local relation between the current and
the eletromotive field
can be obtained if one 
introduces an effective local chemical potential $\mu(\br)$ to
accounts for the density fluctuation\cite{XRS}.  However, equation
(\ref{smunuSCBA})
does not quite recover equation (\ref{classical}). One notices
that in the diffusive term, 
 $\sy^0$ splits into two parts, $\sy^{I,0}$, which appears in the bulk 
diffusion current and $\sy^{II,0}$, which  is proportional to the edges 
current density and shows up only at the reflecting edges
($\sy=\sy^{I}+\sy^{II}$). It was shown
that $\sy^{II}$ is proportional to the
rate at which the total density changes with the magnetic field at
fixed Fermi energy\cite{Pruisken}. In the low 
field limit $\wc\tau_0\ll 1$ ($\wc$ is the cyclotron radius and
$\tau_0$ is the zero field elastic scattering rate), the Landau levels
merge into a continuum and  the density of states is only weakly
dependent on the magnetic field, therefore the $\sy^{II}$ term can
be ignored.  This is not the case in the high field limit with
$\wc\tau_0>1$, when separate Landau bands are formed, and the density of
states oscillates with the field. In
general equation (\ref{classical}) should be replaced by\cite{XRS}
\be
j_{\mu}=\sigma^0_{\mu\nu}E_{\nu}
+[\sx^0 \delta_{\mu\nu} +  \sy^{I,0} \epsilon_{\mu\nu}]\partial_{\nu}\mu/e 
+\sy^{II,0} [\delta(y-W)-\delta(y)]\delta_{\mu,x}\;\mu/e.
\ee
in which we have included  an edge  current  $ I_e=
\sy^{II,0} \mu/e$, which is nothing but the  extra edge current
induced by  an increase in chemical potential.

In the disorder-free leads, $\sy^{I}=0$ and $\sy^{II}=\sy^{lead}=N$ is
quantized, but $\sy^{II}$ 
is not always quantized in the disordered region
\cite{Pruisken}.  	It is apparent that   when the
edge current  in the lead and the
sample are different there has to be an
edge current along the border with the leads amounting to
$I'_e=[\sy^{lead}-\sy^{II,0}]\mu/e$, in order to 
satisfy current conservation at the corners.  
Therefore, for the geometry under consideration the local equation is
further modified to be
\bea
j_{\mu}=&&\sigma^0_{\mu\nu}E_{\nu}
+[\sx^0 \delta_{\mu\nu} +  \sy^{I,0} \epsilon_{\mu\nu}]\partial_{\nu}\mu/e 
+\sy^{II,0}[\delta(y-W)-\delta(y)]\delta_{\mu,x}\;\mu/e\non\\
&&+(\sy^{lead}-\sy^{II,0})\delta_{\mu,y}[\dl(x-L)-\dl(x)]\;\mu.\label{Drudenew}
\eea

To calculate the resistance, there are two approaches. One approach
is to find  the static 
electro-chemical potential  and subsequently the current distribution, by 
requiring $\nabla\times \bf {E}  =-\partial {\bf  B}/\partial t=0 $, in
which case, one can write $\bf {E}(\br)=-\nabla \varphi(\br) $. 
The current conservation condition for the static case is $\nabla \cdot
{\bf j}=0$, which  gives in the 
bulk
\be
\nabla^2 (-\varphi+\mu/e)=0.\label{bulk}
\ee
At the top and bottom edge, we have shown\cite{XRS} that
current conservation leads to the following boundary condition
\be
\sx^0 \partial_y(-\varphi+\mu/e)-\sy^{0}\partial_x 
(-\varphi+\mu/e)=0.\label{BC1} 
\ee 
In other words, the eletromotive field $\cal{E}$ has to make an 
angle $\theta=\tan^{-1} \sy^0/\sx^0$ with the edge along 
$\hat{x}$-direction: 
$$
\frac{{\cal E}_y}{{\cal E}_x}=\frac{\sy^0}{\sx^0}.
$$  
At the left and right border with the leads, 
one can derive the following boundary condition by constructing a surface
that encloses the edge current $I_e'$ and imposing current conservation: 
\be
j_{x}-\partial_y I_e'(y)=j_{x}^{lead}.\label{border}
\ee 
In the leads where $\sx=0$, one can
write down the following equation for 
the local current density\cite{Thouless}: 
\be
j_{\mu}^{lead}=
\sy^{lead}\epsilon_{\mu\nu}E_{\nu}+\sy^{lead}\dl_{\mu,x}
[\dl(y-W)-\dl(y)]\mu/e.\label{lead}
\ee
Combining (\ref{Drudenew}),(\ref{border}) (\ref{lead}), we get
\be
\sx^0 \partial_x(-\varphi+\mu/e) + [\sy^0-\sy^{leads}]
\partial_y(-\varphi+\mu/e)=0, \label{BC2} 
\ee
i. e., $\cal {E}$ has to make an angle 
$\theta'=\tan^{-1}(\sy^0-\sy^{lead})/\sx^0$ with the border with the leads 
(along $-\hat{y}$-direction ):
$$
\frac{{\cal E}_x}{{\cal E}_y}=-\frac{\sy^0-\sy^{lead}}{\sx^0}.$$ 
Thus the problem of finding the resistance 
to leading order of $1/\sx^0$ is that of solving the Laplacian equation
for the electrical 
chemical potential with two tilted boundary conditions. Notice that as 
far as the conductance or resistance is concerned, one can arrive at the 
correct equations by using the simpler but incorrect local relation
(\ref{classical}).  

Alternatively, one can follow the   Landauer-Buttiker approach, i. e., 
one finds the transmission coefficient. This also
requires the computation of the bilocal-conductivity tensor.
 For our
simple geometry,  
\be
T=\int_{S_1} dy_1\int_{S_2} dy_2 \sigma_{xx}(\br_1,\br_2)
\ee
where $S_1$ and $S_2$ are two  the cross-sections  
at the left and right end of the scattering region. Making use of the SCBA
expression (\ref{smunuSCBA}) for
$\sigma_{\mu\nu}(\br,\brp)$ , we get
\bea
T&=&-\frac{1}{\sx^0}\int_0^W dy\int_0^W
dy'(\sx^0\partial_x+\sy^0\partial_y)(\sx^0\partial'_x-\sy^0\partial'_y)
d(\br,\brp)\left|_{x=0,x'=L}\right.
\non\\
&=&\frac{(\sy^{0})^2}{\sx^0}
[d(L,W;0,W)+d(0,0;L,0)-d(L,W; 0,0)-d(L,0; 0,W)].
\eea
In reference\cite{XRS}, we have shown that the diffusion propagator
satisfies  boundary condition  (\ref{BC1}) 
at the reflecting edges, with $d$ replacing $-\varphi+\mu/e$. One can
easily check that
it also satisfies
boundary condition of the form (\ref{BC2}) at borders with the leads.  
The above expression can be simplified if one takes into
account  the additional detailed boundary condition that the returning
probability via the incoming links is zero, the coarse-grained
diffusive version of which is $d(L,0 ;\br)=d(0,W;\br)=0$, and
the fact that transmission probability of left-going and right-going edge
states are
the same, i. e., $d(L,W;0,W)=d(0,0;L,0)$.  We get
\be
T=2\frac{(\sy^{0})^2}{\sx^0}d(L,W;0,W)
\ee 
Defining $\phi(\br)=2\textstyle{\frac{(\sy^{lead})^2}{\sx^0}}d(\br;0,W)$,
 $\phi$ satisfies  (\ref{bulk}), (\ref{BC1}), (\ref{BC2}) and in addition
the
following initial condition $\phi(L,0)=0$, and $\phi(0,W)=N$, since in the
limit $L\rightarrow 0$, $T\rightarrow N$.  Thus within the diffusive limit
the Landauer-Buttiker approach and the current distribution approach  
boil down to the same mathematical problem. Since the problem 
contains non-self-adjoint boundary conditions, its solution in the general
case requires a special method, which we will discuss in the next section.
However, some
conclusions can be made  before  solving the equation.

Again the fact that the Hall resistance  is quantized can be demonstrated
 using boundary condition (\ref{BC2}) alone.   The current across the
border
with the leads is $I_x=\int_0^W dy (\sx^0 { \cal E}_x+ \sy^0 {\cal
E}_y)$ and the
transverse drop in electro-chemical potential at the borders is
$\Delta\phi_y=\int_0^W dy {\cal E}_y(x,y)_{x=0, L}$. Making use of
the relation between ${\cal E}_x$ and ${\cal E}_y$ 
from (\ref{BC2}), the result 
$R_{xy}=\textstyle{\frac{1}{\sy^{lead}}}$ is recovered.

The next conclusion follows from  the two boundary
conditions. Since the field $\cal{E}$ at the reflecting edges and at the
borders with the leads is required to point at two specific directions,  
the  field and current distribution is uniform only when the two
directions coincide, i.
e. when the bare conductivities satisfies the following constraint:
\be
(\sx^0)^2+\sy^0(\sy^0-\sigma^{lead})=0,\label{constraint}
\ee
which is equivalent to 
$$\rho_{xy}=\frac{\sy^0}{(\sx^0)^2+(\sx^0)^2}=\frac{1}{\sy^{lead}}. $$
For $\sy^{lead}=1$, the above gives the same semi-circle constraint
(\ref{semicircle}) of Ruzin et al.
Under condition (\ref{constraint}), the field distributions takes the
simple
form:  
\be
\phi_0(x,y)= - E_0( x + \frac{\sy^0}{\sx^0}  y),\;\;
{\bf \cal E}_0(x,y) = E_0( \hat{x} + \frac{\sy^0}{\sx^0}
\hat{y}),\label{uniform}
\ee
where $E_0$ is a constant.  The
total current, the transverse and longitudinal  electro-chemical 
potential differences can be easily calculated.  
Indeed $R_{xx,SCBA}=\frac{L}{W}\rho_{xx}$.

The $\sx^0$ and $\sy^0$ of the  Chalker-Coddington
model satisfies the above.  However such bare values  were
computed by forcefully leaving out the quantum interference of the
original model\cite{XRS}. Since $\sx^0<1$, the model has no perturbative
regime, i. e., there does not exist a length scale within which these
bare values are the leading contributions. 
For the Gaussian
white noise potential model $\sx^0=(2N-1)\pi^{-1}\sin^2 \alpha $ and
$\sy^0=N-1+\alpha/\pi$, where $ \alpha $ is a function of the Fermi
energy (we do not need its expression here, see \cite{Pruisken}), they do 
not satisfy the constraint. In this case the field and current 
distribution is not  uniform, $R_{xx}$ is not
proportional to $\rho_{xx}$ and $R_{xy}$ is not  equal to
$\rho_{xy}$.

As we mentioned before the geometry we consider (figure 2a) do not
correspond to those
used in the
experiments, we would like to know how $R_{xx}$ and $R_{xy}$
change if we move the voltage probes into the scattering region as shown
in figure 2b.  
Assuming that the voltage probes are very small in dimensions (much
smaller
than the sample length $L$ and width $W$),  we can  treat the probes as  
perturbations in boundary conditions and  get a rough picture  
by looking at the field distribution in the main strip in the absence of
these probes.  If condition (\ref{constraint}) is met and the field and
current
distribution is uniform, the
electro-chemical potential
drop transverse to the current is the same everywhere along the
strip and the longitudinal potential difference is in proportion with
the
distance between the probes. Such systems  have a quantized
Hall resistivity, as it is the case with the
Chalker-Coddington network with no phase coherence or a random version of
it as discussed by\cite{Shimshoni}.  In the general case
when the field is not uniform, 
the Hall resistance departs from the quantized value as we move the
voltage probes to the interior of the sample.  However if the Hall bar is
long and narrow ($L\gg W$), the field distribution in the interior far
from the ends takes
the form  of equation (\ref{uniform}), forced by boundary condition 
(\ref{BC1}). If the probes are placed in the
region where the field is nearly uniform, $R_{xx}$ and $R_{xy}$ are  
roughly proportional to the intrinsic $\rho^0_{xx}$ and $\rho^0_{xy}$. In
this way
the effect of the perfect leads is removed.

\section{Analytical solution of the boundary problem}

To solve the boundary problem we  use the method of 
conformal mapping,  which was first used  by Girvin and 
Rendell\cite{Girvin} to find  the two-probe conductance in the case of
absorbing leads. The idea is as following. Let us consider a
parallelogram in the $z'$-plane ($z'=x'+iy'$), with its top and bottom
edge parallel to $x'$-axis and    its left and right side 
tilted from the vertical direction by $\delta \theta=\theta-\theta'+90$.
In
this geometry, the uniform field distribution of (\ref{uniform}) satisfies
both boundary conditions. The solution for the rectangle geometry in
the $z$-plane ($z=x+iy$) can be
found if one finds  a conformal mapping $z'\rightarrow z$ which
transforms the
parallelogram to the 
rectangle.   We start by writing down the complex potential $\phi(z')$
which can give the correct physical field ${\cal E}_0 (x',y')$. 
The field are related to the complex potential through  
$$
-\frac{d \phi}{d z}={\cal E}_y+i{\cal E}_x.
$$ 
It is easy to see that the following complex potential
\be
\phi(z')=-iE_0 e^{-i\theta} z'.
\ee
gives $\bf{ \cal E}_0(x',y')$ in the $z'$-plane.
The conformal transformation from $z$-plane to the $z'$-plane rotates the
angle 
at $x=0,L $ by $\dl \theta$, i., e.,  the rotation defined by
\be
\frac{d z'}{d z}=e^{f(z)}
\ee
has the boundary conditions
\be
Im [f(z)] =  \delta \theta, \mbox{for}\; x=0,\;L; \;\; Im[f(z)]=0 
\;\;\mbox{for}\;\;y=0, W.
 \ee
The transformation has been  worked out by Girvin and Rendell
to be
\be
f(z)=\sum_{n=odd}\frac{4\dl\theta}{n\pi}\{\sinh[n\pi (z-L/2)/W]/\cosh(n\pi
L/2W)\}.
\ee
The potential drops are:
\be
\Delta\phi_x=-E_0\cos(\theta) \int_{0}^{L} dx \exp{\left[\sum_{n=odd} 
\frac{4\dl\theta}{n\pi}\sinh[n\pi (x-L/2)/W]/\cosh(n\pi
L/2W)\right]},\label{phix}
\ee
\be
\Delta \phi_y=-E_0 sin(\theta-\delta \theta) \int_{0}^{W} dy 
\exp{\left[\sum_{n=odd} 
\frac{4\dl \theta}{n\pi}\tanh(n\pi L/2W)\cos(n\pi
y/W)\right]},\label{phiy} 
\ee
The transmission coefficient is
\be
T=\phi(L,W)=N\frac{\Delta\phi_y}{\Delta \phi_x+\Delta \phi_y}
\ee
In the limit $W=L$, one can show that the two integrand in (\ref{phix})
and (\ref{phiy}) are 
equal, the expression simplifies to be
\be
T_{L=W}=N \frac{\sin(\theta-\dl\theta)}{\cos\theta+\sin(\theta-\dl\theta)}
=N\frac{\cos(\theta')}{\cos(\theta')-\cos\theta}.
\ee
The resistance $R_{xx}$ can be obtained by either plugging the above
into the Landauer-Buttiker expression (\ref{Rxx}) or by
calculating the total current from the field distribution, then dividing
$\Delta\phi_{x}$ by it.

\section{Conclusions and discussions}

Prompted by the experimental evidence of duality  near quantum Hall
transitions, we probe the implication of duality 
in the Chalker-Coddington model within the fermionic non-interacting
picture. For finite systems with boundaries, a duality relation holds for
the edge state transmission coefficient, however, only for a class
of systems with reflection symmetry. For such systems  duality
leads to measurable inverse relations of the longitudinal resistance 
of two states at  opposite sides of a quantum Hall transition. 
In general situation (for arbitrary geometry) the duality translates to a
diluted version $\sy(\sy^0)+\sy(n-\sy^0)=n$. Our key difference with
previous authors on 
the subject\cite{Shahar2} is that we think that the Landauer Buttiker
argument constructed for resistance in symmetric  geometry can not be
expected to hold in other geometries, such as a Hall bar, since
our calculations shows that resistance is not necessarily
resistivity. If experiment of Shahar et al is correct and the reversal
of role of longitudinal voltage and current is indeed observed in the
quantum scaling regime, it points to the possibility that the renormalized
$\sx$ and $\sy$ satisfy the constraint (\ref{semicircle}) first suggested
by the numerical work of Ruzin et al. This has so far not been understood within
any theoretical framework.

Our work unravels the mystery surrounding the Landauer-Buttiker
argument. We performed a perturbative calculation for the 
edge state transmission coefficient in terms of
the conductivities and analyze the current distribution. The presence
of the 2D perfect leads gives rise to a tilted boundary condition for
diffusion, which is similar and dual to the one at the reflecting edges,
(dual in the sense that one boundary condition becomes the other as
$\sy^0\rightarrow n-\sy^0$).
Only when $\sx$ and $\sy$ satisfies the
semi-circle constraint (equivalent to the quantization of $\rho_{xy}$),
the current distribution is uniform and the resistance $R_{xx,xy}$ is in
proportion to the resistivity $\rho_{xx,xy}$. 
Our phenomenological treatment 
is based on our knowledge of  the bilocal
conductivity $\sigma_{\mu\nu}(\br,\brp)$ in the perturbative regime.  To
access the finite-size scaling of the resistance in the
non-perturbative critical regime, one needs  to study the form of 
$\sigma_{\mu\nu}(\br,\brp)$ afresh.

It is worth noting that we have considered the consequences of duality
only within  the zero-temperature,  disordered and phase-coherent  
models for the quantum Hall transitions, while experiments are 
done at finite temperature where the 
electron-electron interaction (or the interactions between 
quasi-particles in the 
case of the fractional quantum Hall effect) must be 
considered.   At a preliminary level interaction  
gives rise to a finite 
phase-coherent length which acts as 
a size cut-off in the renormalization of the conductivities\cite{Pruisken} 
when the elastic scatterings dominate over 
the phase-randomnizing inelastic scatterings. 
However, whether or not 
interactions are relevant at the quantum Hall transitions is 
an open question\cite{LW2} and there are indications that the non-interacting
picture for the transitions do not embrace all experimental facts 
(see \ref{Subir} and the references therein).
There have been interesting works which reveal non-trivial finite-temperature 
transport properties  near critical transitions where interactions play 
key roles\cite{Subir,Kedar}. An explanation for the observed duality may lie
in the understanding of the full problem with both interaction and disorder.

\acknowledgements

This work is based in part on my previous collaboration 
with N. Read and A. D. Stone. I  thank  A. Nersesyan for helpful
comments, Z. B. Su for clarifying conversations, and J. Moreno for
a critical reading of the manuscript.


\begin{figure}

\caption{Edge states scattering through a disordered region.}
\label{}
\end{figure}
\begin{figure}

\caption{The Hall bar geometry. a: fictitious version with 
the voltage probes set outside the disordered region. b: more realistic
version with the voltage probes set in the interior of the sample.}
\label{}
\end{figure}

\begin{figure}

\caption{Duality in the Chalker-Coddington network model for the quantum 
Hall effect. At fixed Fermi energy $E_f$, the system can be viewed  as
a network built of counter-clockwise orbits enclosing the shaded areas
with tunneling probability $1-T_0(E_f)$,
or equivalently, a network built of clockwise orbits enclosing the white
regions with tunneling probability $T_0(E_f)$. The phase at $T_0(E_f)$ can
be mapped onto the phase at
$T_0(E'_f)=1-T_0(E_f)$ upon reflection with respect to one of the diagonal 
axis.} 
\label{}
\end{figure}
\newpage
\psfig{figure=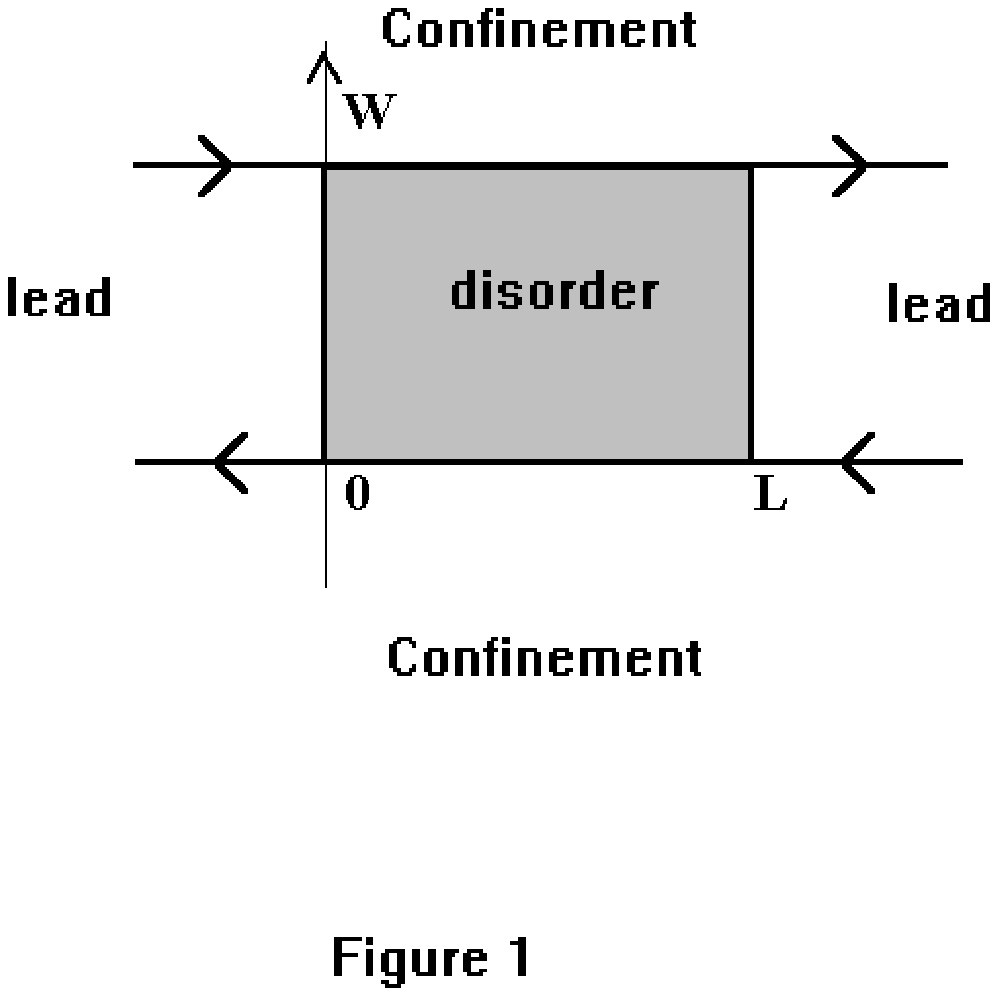}
\newpage
\psfig{figure=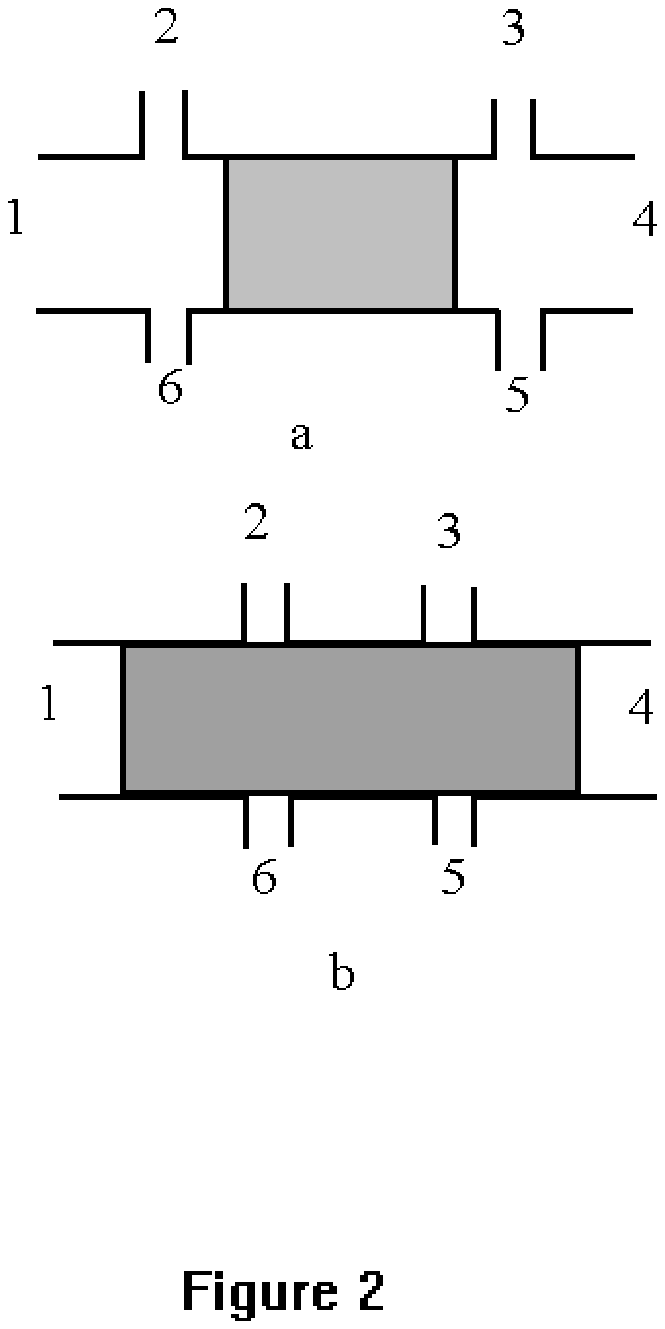}
\newpage
\psfig{figure=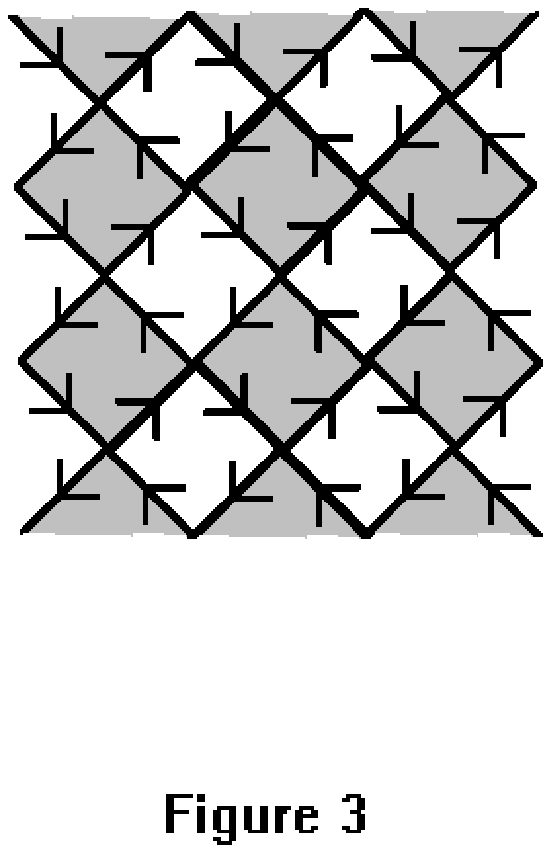}
\end{document}